# Timing-Fragility–Aware Selective Hardening of RISC-V Soft Processors on SRAM-Based FPGAs

Mostafa Darvishi, *Senior Member, IEEE*

***Abstract*—Selective hardening is widely employed to improve the reliability of FPGA-based soft processors while limiting the overhead of full redundancy. However, existing approaches primarily rely on architectural criticality or functional fault analysis, overlooking the impact of routing-dependent timing sensitivity on processor robustness. This paper introduces a timing-fragility–aware selective hardening methodology for RISC-V soft processors implemented on SRAM-based FPGAs.**

**Building on recent advances in in-situ timing observability, the proposed approach quantifies the statistical timing sensitivity of pipeline components under controlled routing perturbations and uses this information to guide hardening decisions. Experimental results on a RISC-V processor implemented on a commercial FPGA platform show that components exhibiting higher timing fragility also demonstrate increased vulnerability to routing-induced delay effects. Leveraging this correlation, the proposed selective hardening strategy achieves robustness comparable to full hardening while significantly reducing area and timing overhead.**

**These results demonstrate that timing fragility provides a practical and effective metric for reliability-aware design optimization in FPGA-based processor architectures.**

## I. INTRODUCTION

FPGA-based soft processors, particularly RISC-V implementations, are increasingly adopted in embedded, aerospace, and adaptive computing systems due to their flexibility and rapid design turnaround. In such environments, reliability is a critical concern, motivating extensive research into fault-tolerant processor architectures and selective hardening techniques [1], [2].

Prior works have shown that selective hardening such as partial triple modular redundancy (TMR) or duplication of critical modules, can significantly improve robustness while avoiding the prohibitive overhead of full redundancy [3], [4]. However, existing selective hardening methodologies typically base their decisions on architectural importance, execution frequency, or functional fault coverage. While effective at a high level, these approaches implicitly assume uniform physical behavior across the processor fabric.

In practice, the timing behavior of FPGA-based processors is strongly influenced by routing topology, interconnect parasitics, and power distribution effects [5], [6]. Recent studies have demonstrated that timing degradation often precedes functional failure and that routing-induced delay perturbations can significantly affect processor robustness [7], [8]. Nevertheless, timing sensitivity has not yet been incorporated as a first-class criterion in selective hardening decisions.

This paper addresses this gap by introducing a timing-fragility–aware selective hardening framework for RISC-V soft processors. Building on prior work that established an in-situ timing diagnosis architecture for SRAM-based FPGAs, we leverage statistical timing observability to identify pipeline components that are particularly sensitive to routing-induced delay effects. Rather than treating all architecturally critical modules equally, the proposed approach prioritizes hardening based on measured timing fragility.

Using an FPGA-based RISC-V implementation, we show that timing fragility correlates strongly with vulnerability to routing perturbations and that selective hardening guided by this metric can achieve robustness comparable to full hardening with substantially lower overhead. These results demonstrate that timing-aware analysis provides a powerful and previously underutilized dimension for reliability-driven design optimization in FPGA soft processors.

## II. BACKGROUND AND RELATED WORKS

SRAM-based FPGAs have become a widely adopted platform for implementing soft processors in safety-critical, space-constrained, and adaptive computing systems due to their flexibility and rapid deployment capabilities [9]-[10]. However, the use of configuration memory to define both logic and routing resources introduces intrinsic reliability challenges. Configuration bits stored in SRAM cells are susceptible to disturbances arising from voltage fluctuations, aging, and environmental effects, which can manifest as functional faults or performance degradation in deployed designs [11], [12].

Soft processors implemented on FPGAs, such as RISC-V cores, are particularly sensitive to these effects because their correctness depends not only on logical functionality but also on meeting strict timing constraints across deeply pipelined datapaths [13]-[18]. Unlike ASIC implementations, FPGA-based processors rely heavily on programmable routing resources whose delay characteristics can vary significantly across the fabric [19]. As a result, reliability threats in FPGA soft processors often emerge as *timing failures* rather than explicit logical errors [6], [19]-[21].

Traditional fault-tolerance techniques for FPGA-based processors have primarily focused on functional correctness, often assuming that timing margins remain intact [22]-[24].

Mostafa Darvishi is with Electrical Engineering Department of École de technologie supérieure (ÉTS), Montreal, Canada. He is also VP of Engineering at Evolution Optiks R&D Inc. (e-mail: darvishi@ieee.org).

However, growing evidence suggests that timing degradation—especially in routing-dominated paths—represents a critical and underexplored failure mode in modern FPGA implementations [25]-[29].

To address the high area and power overhead of full duplication or triple modular redundancy (TMR), prior work [30] has explored *selective hardening* strategies that protect only a subset of processor components . These approaches typically rely on architectural or functional analysis to identify critical modules, such as register files, control logic, or pipeline stages, that disproportionately affect system correctness.

The works presented in [31] and [32] represent state-of-the-art examples of this paradigm. In these studies, selective hardening is applied to RISC-V soft processors by identifying architecturally critical elements and selectively replicating or protecting them to improve reliability. The selection criteria are largely driven by functional vulnerability metrics, architectural role, or instruction-level impact, enabling substantial overhead reduction compared to blanket redundancy.

While effective at mitigating certain classes of faults, these approaches implicitly treat all instances of a given architectural structure as equivalent. That is, protection decisions are made at the level of processor components (e.g., pipeline registers, execution units) without accounting for the *physical implementation context* of those components within the FPGA fabric.

A key limitation of existing selective hardening techniques is the absence of physical timing awareness in the protection strategy. In FPGA-based implementations, two logically identical modules may exhibit vastly different timing behavior depending on their placement, routing topology, and interaction with surrounding interconnect resources. Consequently, architectural criticality does not necessarily correlate with *timing fragility*.

The prior works do not explicitly measure or exploit variations in routing-induced delay sensitivity across the processor. As a result, selectively hardened designs may overprotect timing-robust regions while leaving timing-fragile paths exposed. This mismatch can lead to inefficient use of redundancy resources and limit the achievable reliability gains under realistic operating conditions.

Furthermore, most existing studies rely on static analysis, fault injection at the functional level, or architectural modeling to evaluate reliability. These methods provide limited visibility into how timing behavior evolves across the FPGA fabric during operation and how local routing perturbations translate into observable delay shifts.

Recent advances in in-situ timing measurement techniques have demonstrated the feasibility of observing routing-level timing behavior directly within FPGA fabrics. Such approaches enable fine-grained characterization of delay distributions, variability, and spatial correlation across different regions of a design. These measurements reveal that timing degradation mechanisms, such as routing parasitics and configuration-induced perturbations, exhibit strong locality and nonuniformity.

This observation motivates a new class of selective hardening strategies that move beyond purely architectural considerations and instead incorporate *measured timing fragility* as a first-class selection criterion. By identifying which portions of a processor are most sensitive to routing-induced delay variation, redundancy and hardening resources can be targeted where they are most impactful.

In contrast to prior work, a timing-fragility–aware approach enables protection decisions to be guided by empirical physical behavior rather than abstract architectural importance alone. This paradigm offers the potential to achieve higher reliability efficiency, reducing overhead while directly addressing the dominant sources of timing failure in FPGA-based soft processors.

Building on the selective hardening concepts introduced in [31] and [32] and leveraging in-situ timing observability mechanisms similar in spirit to those explored in recent timing diagnosis studies, this paper proposes a timing-fragility–aware selective hardening framework for RISC-V soft processors implemented on SRAM-based FPGAs.

Unlike prior approaches, the proposed method explicitly incorporates spatially resolved timing measurements to guide hardening decisions. By correlating timing fragility with architectural structures, the framework enables selective protection of only those processor regions that are both architecturally important and physically vulnerable. This integration of physical timing behavior with architectural hardening represents the key departure from existing work and forms the foundation for the contributions presented in the remainder of this paper.

## III. TIMING FRAGILITY AND MOTIVATION FOR SELECTIVE HARDENING

Timing reliability in SRAM-based FPGAs is governed not only by nominal delay margins but also by the sensitivity of routed signal paths to perturbations in the programmable interconnect fabric. Conventional static timing analysis (STA) ensures that setup and hold constraints are satisfied under worst-case assumptions, yet it provides limited insight into how close individual portions of a design operate to their effective timing margins under variability, noise, or configuration-induced perturbations. In FPGA-based soft processors, where logic functionality and routing resources are tightly interwoven, this limitation becomes particularly pronounced.

In practice, timing vulnerability is not uniformly distributed across a processor pipeline. Even when all pipeline stages satisfy nominal timing constraints, their susceptibility to routing-induced delay variation can differ substantially due to differences in routing depth, fan-out, and switch-matrix traversal. This section establishes timing fragility as a measurable and architecturally relevant property and demonstrates its role as a key motivation for selective hardening.

Fig. 1 illustrates the conceptual framework used to localize timing fragility within an FPGA-implemented RISC-V processor. The processor pipeline is decomposed into its constituent stages—Instruction Fetch (IF), Instruction Decode (ID), Execute (EX), Memory Access (MEM), and Write-Back (WB), each mapped onto distinct regions of the FPGA fabric. Signals associated with these stages traverse programmable routing structures composed of switch matrices and interconnect segments whose electrical characteristics vary spatially across the device. To observe timing behavior without disturbing functional execution, non-intrusive timing probes are attached to representative routing paths associated with each pipeline stage. These probes extract statistical timing information directly from routed signals, enabling stage-resolved characterization while the processor remains fully operational.

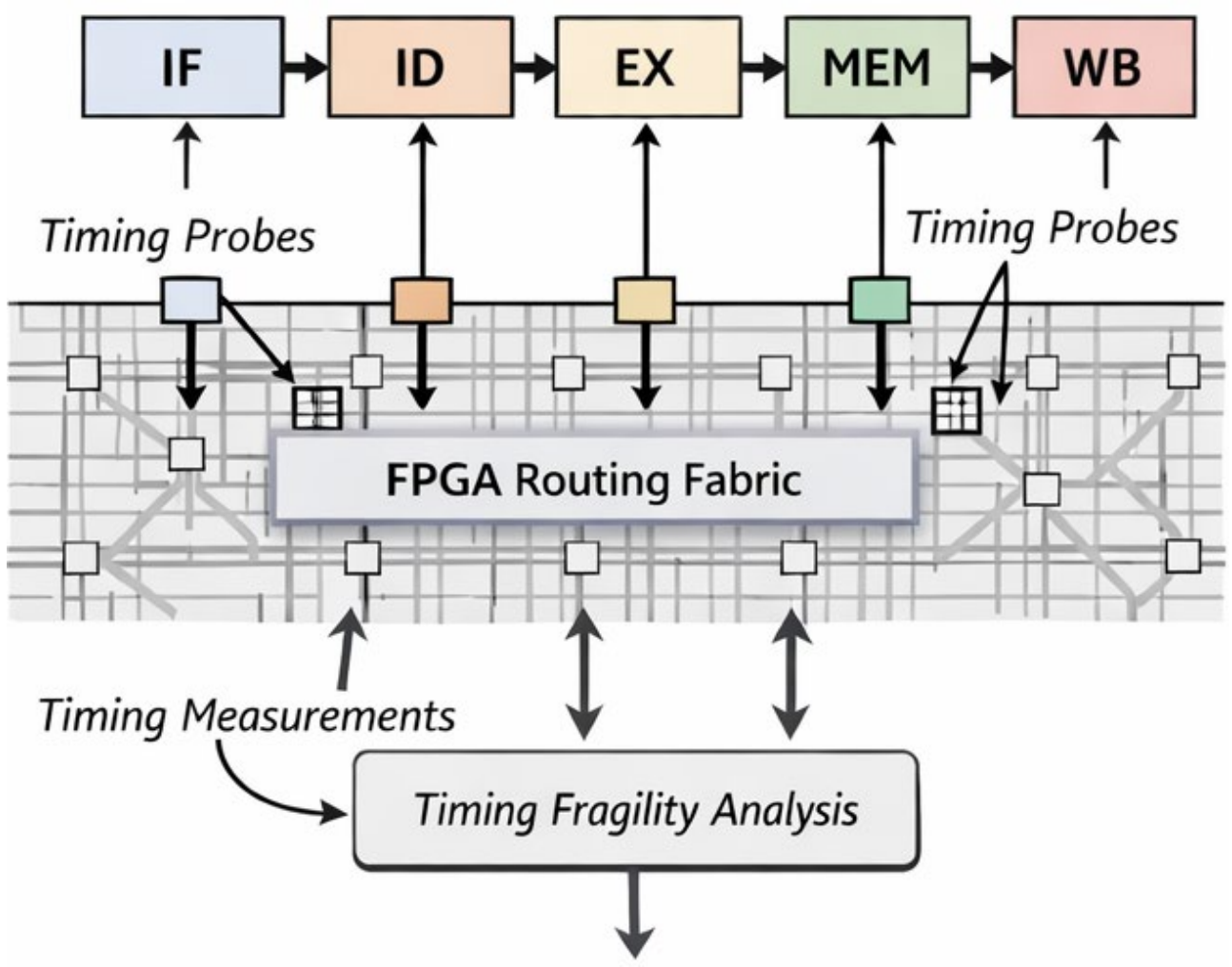

Fig. 1. Pipeline-stage–resolved timing fragility mapping in an FPGA-based RISC-V processor. Distributed timing probes observe routing-level delay behavior associated with each pipeline stage, enabling localized identification of timing-robust and timing-fragile regions.

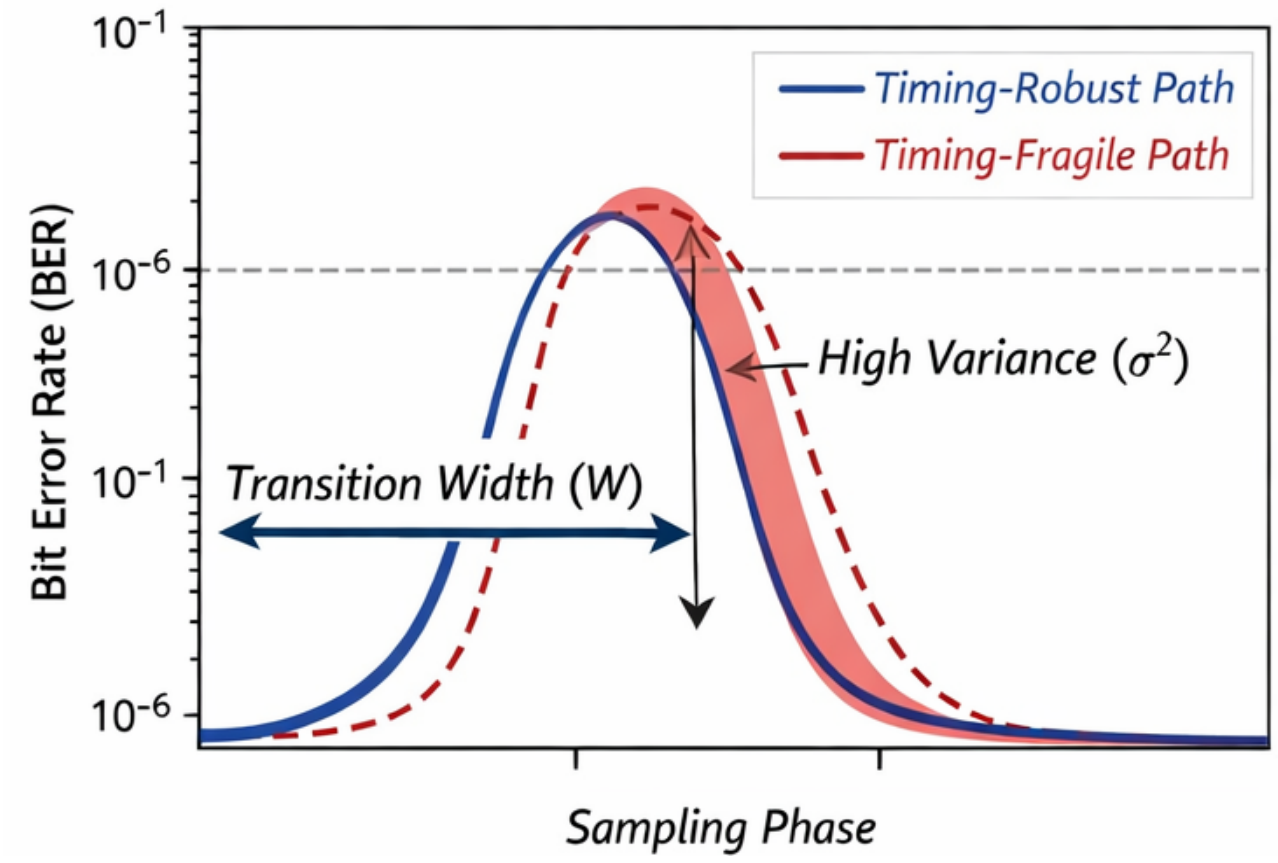

Fig. 2. BER versus sampling-phase characteristics for representative timing-robust and timing-fragile routing paths. Fragile paths exhibit wider transition regions and increased variance, indicating heightened sensitivity to delay perturbations despite satisfying nominal timing constraints.

While Fig. 1 highlights where timing fragility resides architecturally, understanding how fragility manifests requires examining the statistical behavior of signal transitions under marginal timing conditions. To this end, timing behavior is characterized using phase-swept sampling, which produces a probabilistic view of signal arrival times rather than a binary pass/fail outcome.

Fig. 2 presents representative bit-error-rate (BER) versus sampling-phase characteristics for two routed paths extracted from the processor: one exhibiting timing-robust behavior and one exhibiting timing-fragile behavior. Both paths satisfy nominal STA constraints; however, their statistical profiles differ markedly. The timing-robust path exhibits a narrow transition region in which BER changes sharply with sampling phase, indicating a well-defined transition point and limited sensitivity to small delay variations. In contrast, the timing-fragile path exhibits a substantially wider transition region and increased variance, reflecting heightened sensitivity to routing-induced delay perturbations arising from interconnect parasitics, power-distribution noise, or local configuration effects.

The transition width and variability illustrated in Fig. 2 form the basis of the quantitative timing fragility metrics used in this work. Specifically, the width of the BER transition region provides a direct statistical indicator of how vulnerable a path is to delay perturbations, while the variability of the transition location across repeated measurements captures the stability of that path under marginal conditions. These metrics expose timing vulnerability that is not reflected in nominal slack values.

The implications of these statistical differences across the processor pipeline are summarized in Table I, which reports representative timing fragility metrics extracted from the implemented RISC-V processor and is used here to motivate selective hardening decisions. A comprehensive experimental evaluation is presented later in Section VI. All stages satisfy nominal static timing constraints, with positive slack margins ranging from 0.33ns to 0.46ns. However, statistical measurements reveal pronounced non-uniformity in timing robustness.

Table I shows that the Execute (EX) and Memory Access (MEM) stages exhibit substantially wider BER transition regions and higher timing variability than the Instruction Fetch (IF) and Instruction Decode (ID) stages, despite comparable nominal slack. These differences are attributed to increased routing depth and fan-out in later pipeline stages, which amplify sensitivity to routing-level delay perturbations. Conversely, stages with simpler routing structures demonstrate greater intrinsic robustness even when static timing margins are similar.

Based on these observations, pipeline stages are classified into *low*-, *moderate*-, and *high-fragility* categories, as indicated in Table I. This classification is derived directly from measured statistical behavior rather than heuristic assumptions and demonstrates that static timing slack alone is insufficient for identifying timing-vulnerable regions under realistic operating conditions. The results motivate a selective hardening strategy in which mitigation resources are applied preferentially to statistically fragile stages, reducing overhead while addressing the dominant sources of timing vulnerability.

Table I. Pipeline-stage timing characteristics and statistical fragility metrics.

| Pipeline Stage | Nominal Slack (ns) | Mean Delay μ (φ) | Transition Width Δφ (φ) | Std. Dev. σφ (φ) | Fragility |
|---|---|---|---|---|---|
| **IF** | 0.46 | 0.48 | 0.015 | 0.003 | Low |
| **ID** | 0.41 | 0.51 | 0.019 | 0.005 | Low |
| **EX** | 0.33 | 0.57 | 0.041 | 0.014 | High |
| **MEM** | 0.36 | 0.60 | 0.046 | 0.017 | High |
| **WB** | 0.43 | 0.53 | 0.022 | 0.006 | Moderate |

In summary, this section establishes timing fragility as a measurable, pipeline-dependent property in FPGA-based RISC-V processors. By combining stage-resolved localization (Fig. 1), statistical path characterization (Fig. 2), and quantitative comparison across pipeline stages (Table I), the section provides a rigorous foundation for the selective hardening framework introduced in the next section.

## IV. SELECTIVE HARDENING ARCHITECTURE

This section presents the proposed selective hardening architecture for FPGA-based RISC-V soft processors. The architecture is designed to exploit the timing fragility characterization introduced in Section III, enabling targeted mitigation of routing-induced timing vulnerabilities while minimizing area, power, and performance overhead. Rather than applying uniform redundancy across the processor, the proposed approach selectively hardens only those architectural regions that exhibit statistically significant timing fragility.

The central design principle of the architecture is the decoupling of *architectural importance* from *physical timing vulnerability*. While prior selective hardening approaches focus primarily on architectural role, such as control logic or register files, the proposed architecture introduces timing fragility as an additional, physically grounded criterion. This enables protection resources to be allocated where they are both architecturally relevant and physically vulnerable.

### *A. Hardening Granularity and Target Selection*

To mitigate timing vulnerability while preserving implementation efficiency, this work adopts a *selective hardening strategy* that targets only the most timing-critical portions of the processor pipeline. Rather than uniformly hardening the entire core, which would incur prohibitive area, power, and performance penalties, the proposed approach leverages pipeline-stage–level timing characterization to localize hardening where it yields the highest return.

Fig. 3 illustrates the proposed selective hardening architecture for a RISC-V processor core implemented on an SRAM-based FPGA. The baseline pipeline consists of the instruction fetch (IF), instruction decode (ID), execute (EX), memory access (MEM), and write-back (WB) stages. In the proposed approach, redundancy is selectively applied to the EX and MEM stages, while the remaining stages retain their original, unhardened implementation. This choice is driven by the fact that the EX and MEM stages collectively account for the majority of critical-path depth, routing complexity, and timing sensitivity in typical RISC-V microarchitectures.

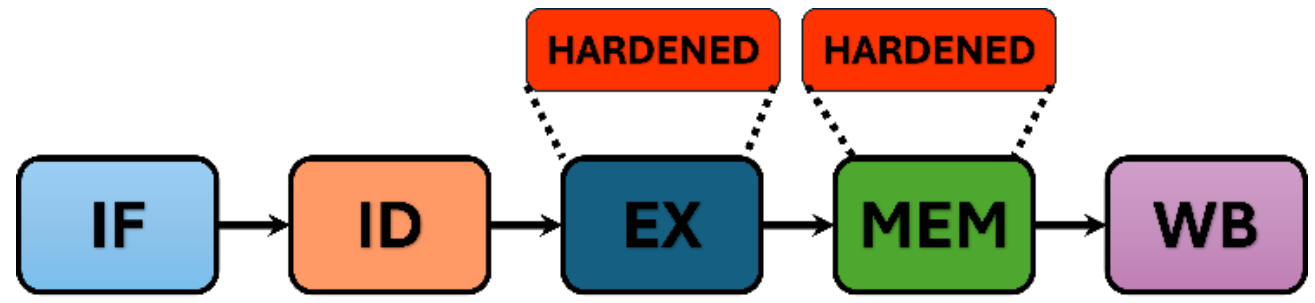

Fig. 3. Selective hardening applied to timing-fragile pipeline stages within a five-stage RISC-V processor. Hardening is selectively introduced at the Execute (EX) and Memory (MEM) stages, which are identified as the dominant contributors to timing criticality and delay variability, while the Instruction Fetch (IF), Instruction Decode (ID), and Write-Back (WB) stages remain unmodified to minimize area and power overhead.

In the proposed architecture shown in Fig. 3, selective hardening of the EX and MEM stages (shown as HARDENED modules) may be realized using either duplication with comparison or triple modular redundancy (TMR), depending on reliability requirements. In both cases, only the targeted pipeline stages are replicated, and lightweight comparison or voting logic is inserted at stage boundaries to detect or mask faults before corrupted data propagates downstream. Importantly, this localized insertion of redundancy avoids introducing global feedback paths or long voter chains that could otherwise exacerbate routing congestion and degrade timing closure.

Compared to full-core hardening approaches, such as those adopted in [31] and [32], where duplication or TMR is applied uniformly across all pipeline stages and control logic, the proposed selective strategy significantly reduces structural overhead while preserving protection where it is most impactful. Full-core hardening provides uniform fault coverage but incurs substantial penalties in logic utilization, routing demand, and power consumption, which can in turn introduce new timing vulnerabilities. By contrast, selective hardening aligns protection granularity with architectural criticality.

To quantify these trade-offs at the architectural level, ` summarizes the expected overhead ranges associated with selective and full hardening strategies. These values reflect typical FPGA synthesis behavior reported in prior works [7], [30], [31], [32] and vendor tool analyses and are scaled according to the fraction of the pipeline protected. When redundancy is confined to the EX and MEM stages, selective duplication is expected to incur approximately 15–25% additional LUT/FF overhead with an 8–18% increase in dynamic power, while selective TMR raises these figures to roughly 25–35% and 15–25%, respectively. In contrast, full duplication and full TMR approaches, such as those employed in [31] and [32], commonly exceed 80% and 180% logic overhead, with corresponding power increases that can approach or exceed 100%, alongside severe routing congestion.

Crucially, these overhead expectations highlight that selective hardening not only reduces resource consumption but also mitigates secondary timing degradation caused by excessive routing complexity. By limiting redundancy to the most vulnerable pipeline stages, the proposed architecture achieves a more favorable balance between reliability, performance, and implementation cost. The detailed implementation of the selective hardening mechanisms and their experimental validation is presented in the subsequent sections.

Table II. Expected overhead ranges for selective versus full hardening strategies in FPGA-based RISC-V cores

| Hardening Strategy | Protected Scope | LUT / FF Overhead | Dynamic Power Overhead | Routing Congestion Impact |
|---|---|---|---|---|
| Selective Duplication | EX + MEM pipeline stages only | 15% – 25% | 8% – 18% | Low |
| Selective TMR | EX + MEM pipeline stages only | 25% – 35% | 15% – 25% | Moderate |
| Full Duplication | Entire core (all stages + control) | 80% – 120% | 40% – 70% | High |
| Full TMR | Entire core (all stages + control) | 180% – 250% | 90% – 150% | Very High |

### B. Selective Hardening Mechanisms and Pipeline Integration

Selective hardening in the proposed architecture is achieved by introducing redundancy only within pipeline stages that are identified as reliability-critical, while leaving the remainder of the pipeline unmodified. As shown in Fig. 4, hardening is selectively applied to the execute (EX) and memory access (MEM) stages, whereas the instruction fetch (IF), instruction decode (ID), and write-back (WB) stages retain their original single-instance implementations. This selective deployment avoids the excessive area and routing overhead associated with full-core hardening techniques.

Two redundancy mechanisms are supported and explicitly illustrated in Fig. 4: duplication with comparison and triple modular redundancy (TMR). In the duplication-based configuration (upper portion of Fig. 4), two identical replicas of the EX stage (EX_A and EX_B) operate in parallel on the same input operands received from the ID stage. Their outputs are evaluated by an EX comparator positioned at the EX stage boundary. Any mismatch between the two replicas asserts an *EX_Error_Flag*, indicating the presence of a fault. An analogous structure is applied to the MEM stage, where MEM_A and MEM_B feed a MEM comparator prior to write-back, generating a *MEM_Error_Flag* upon disagreement.

In the TMR-based configuration (lower portion of Fig. 4), three replicas of the EX stage (EX_1, EX_2, EX_3) and three replicas of the MEM stage (MEM_1, MEM_2, MEM_3) execute concurrently. Majority voters placed at the EX and MEM stage boundaries select the correct output value in the presence of a single faulty replica. In addition to masking faults, the voters also generate stage-level error flags when disagreement is detected among the replicas. As in the duplication case, all redundancy logic is strictly confined within the boundaries of the hardened stages.

Pipeline integration is realized by inserting comparison or voting logic immediately before the pipeline registers that feed subsequent stages, as depicted at the EX→MEM and MEM→WB interfaces in Fig. 4. This placement ensures that erroneous data produced within a hardened stage is either detected or corrected before it can propagate further along the pipeline. Importantly, no global feedback paths or cross-stage dependencies are introduced, preserving the locality of the hardening logic.

Error signaling is handled through dedicated stage-level error flags, explicitly labeled in Fig. 4 as *EX_Error_Flag* and *MEM_Error_Flag*. These flags are routed to an *Error Aggregator*, which consolidates fault indications from multiple hardened stages. The aggregated fault information is then stored in a *Fault Status Register*, providing a centralized and persistent record of detected errors. This error-reporting path is logically separate from the datapath and does not lie on any timing-critical signal path.

The *Pipeline Control Unit* interfaces with the fault status register to enable appropriate system-level responses, such as pipeline stalling, instruction replay, or exception handling, without requiring structural modifications to the core pipeline logic. Because the hardened stages preserve the same input–output interface as their baseline counterparts, existing control mechanisms—including hazard detection, forwarding, and scheduling—remain unchanged.

Selective hardening activation is governed by a *Hardening Configuration Register*, shown in Fig. 4. This register determines which pipeline stages are hardened and whether duplication or TMR is employed. By externalizing hardening control into a configuration register, the architecture supports flexible reliability–performance trade-offs without requiring resynthesis or redesign.

Compared to full-core duplication and TMR approaches reported in [31] and [32], the selective mechanism illustrated in Fig. 4 substantially limits logic replication and interconnect growth. By aligning redundancy with architectural vulnerability, the proposed design achieves improved fault coverage with significantly lower area, power, and timing overhead, making it well suited for FPGA-based RISC-V systems where both reliability and performance are critical.

### C. Hardening Configuration and Overhead Trade-Offs

A central objective of the proposed selective hardening framework is to provide fine-grained configurability that allows reliability mechanisms to be tailored to application requirements and operating conditions while explicitly controlling implementation overhead. Unlike full-core hardening approaches, which apply uniform redundancy regardless of vulnerability distribution, the proposed architecture enables selective activation of protection mechanisms at the level of individual pipeline stages and control boundaries.

Configuration flexibility is achieved through parameterizable hardening control signals that govern (i) which pipeline stages are hardened, (ii) the type of redundancy applied—either duplication with comparison or triple modular redundancy (TMR)—and (iii) the fault-handling policy triggered upon error detection. As illustrated in Fig. 4, these selections are managed through a dedicated hardening configuration register that enables or disables stage replication and selects the appropriate comparison or voting logic. Depending on system requirements, these controls may be statically defined at synthesis time or dynamically programmed at runtime, allowing a single processor instance to operate in multiple reliability modes.

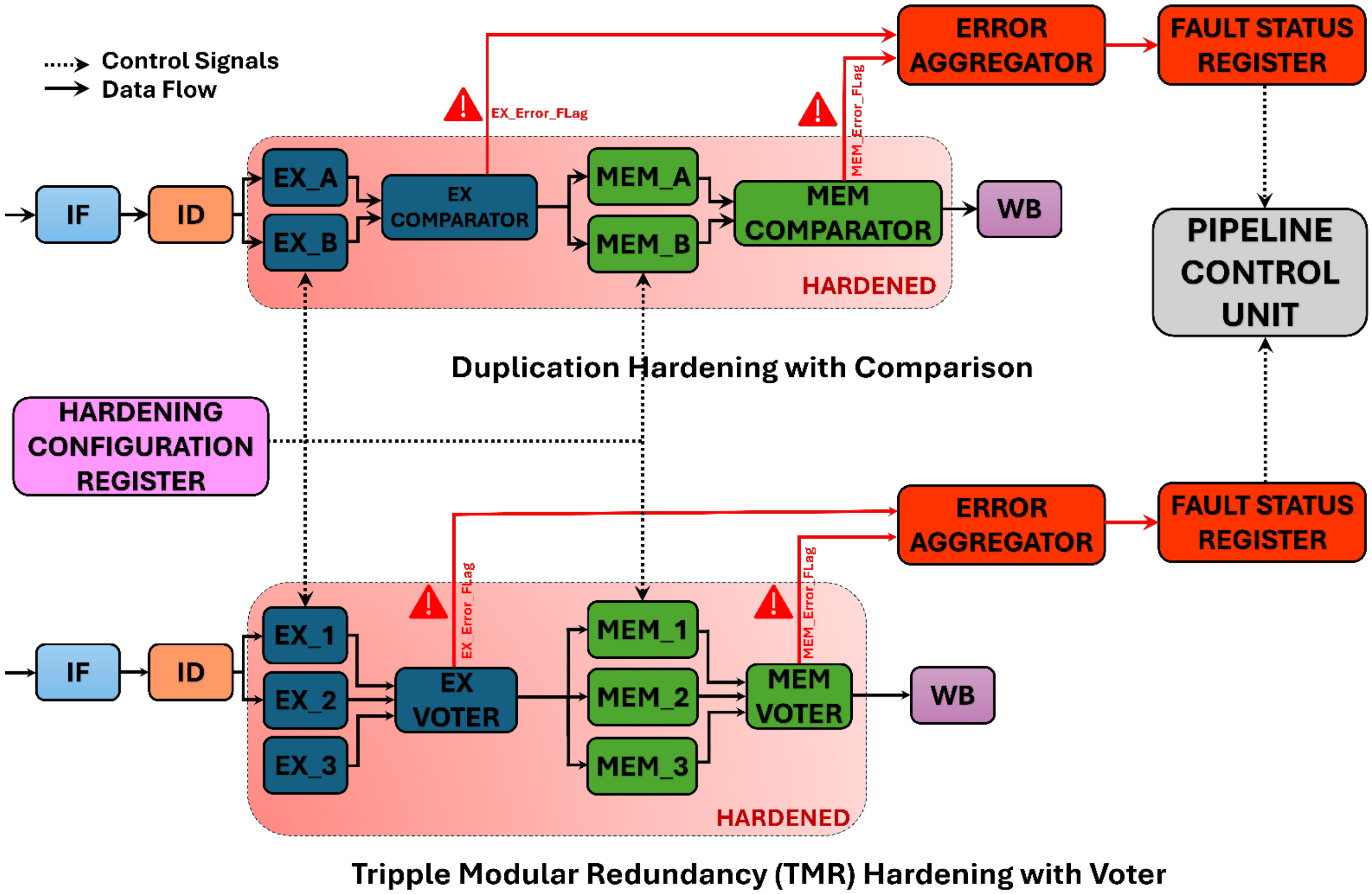

Fig. 4. Selective hardening integration at pipeline stage boundaries. The diagram shows duplication hardening (upper part of figure), and triple modular redundancy (TMR) hardening (lower part of figure) applied to the EX and MEM stages of a RISC-V pipeline. Comparison or voting logic is inserted locally at stage boundaries, while IF, ID, and WB stages remain unhardened.

This configurability directly enables overhead–reliability trade-offs, which are summarized in Table II. Selective duplication incurs moderate area and power overhead by replicating only targeted pipeline stages and inserting lightweight comparison logic at stage boundaries. Selective TMR further increases overhead but provides fault masking capability, preventing single-event faults from propagating beyond the hardened stage. In contrast, full duplication and full TMR replicate the entire pipeline and introduce global comparison or voting logic, resulting in substantially higher area, power, and routing overhead.

Importantly, selective hardening confines redundancy and associated control logic to localized regions of the pipeline, as highlighted by the shaded hardened regions in Fig. 4. This localization reduces the introduction of long interconnect paths and avoids centralized voters or comparators that could otherwise degrade timing closure. As reflected in Table II, selective approaches consistently offer lower overhead than their full-core counterparts while still targeting the pipeline stages most susceptible to timing degradation and transient faults.

From a design perspective, this flexibility enables system architects to align protection levels with application-specific constraints. For example, safety-critical workloads may enable selective TMR on execution and memory stages, while energy-constrained applications may rely on selective duplication or disable hardening entirely during benign operating conditions. This adaptability distinguishes the proposed framework from prior full-core hardening techniques and positions it as a practical solution for reliability-aware RISC-V designs in FPGA-based systems.

### D. Integration with Pipeline Control and Exception Handling

For selective hardening to be practical in a high-performance RISC-V processor, it must integrate seamlessly with existing pipeline control, hazard management, and exception-handling mechanisms. The proposed architecture is explicitly designed to operate within the standard pipeline control framework, ensuring that the introduction of redundancy and error signaling does not disrupt functional correctness or timing determinism.

At the pipeline level, hardened stages operate synchronously with their non-hardened counterparts and preserve original stage boundaries. When duplication with comparison is employed, comparator logic is placed at the output of the protected stage and generates an error flag only when a mismatch between replicas is detected. In the case of selective TMR, the voter produces a single masked output while optionally asserting an error indication when disagreement among replicas is observed. In both cases, hardened stages present a single, timing-compatible output to downstream logic, allowing them to interface transparently with unmodified pipeline stages.

As shown in Fig. 4, error flags generated at the execute and memory stages are propagated through dedicated control paths to an error aggregation unit and fault status registers, rather than being fed back into the datapath. These status signals are sampled by the pipeline control unit at well-defined clock boundaries, ensuring that fault detection does not introduce asynchronous behavior or violate pipeline timing assumptions. Upon detection of an error, the control unit may invoke standard recovery actions—such as stalling, flushing, or replaying the affected instruction—depending on the configured fault-handling policy.

Crucially, the proposed framework does not require modifications to the architectural state or instruction set. Error handling leverages existing exception and interrupt mechanisms, allowing detected faults to be surfaced to software as recoverable exceptions, machine-check events, or logged diagnostic signals. This preserves compatibility with existing RISC-V software stacks and avoids the need for custom fault-aware instruction semantics.

Selective hardening also integrates naturally with pipeline hazard management. Because redundancy is confined to individual stages and does not introduce global feedback loops or long dependency chains, existing hazard detection and forwarding logic continue to operate correctly without modification. Comparator and voter logic are placed outside the critical paths of dependency resolution networks, minimizing their impact on pipeline timing.

### *E. Design Scalability and Portability*

An important objective of the proposed selective hardening framework is to support scalability across increasingly complex processor designs while remaining portable across implementation platforms. The architecture is therefore structured to scale with pipeline depth, issue width, and core count without introducing centralized bottlenecks or rigid design assumptions.

At the single-core level, scalability is achieved by confining redundancy and error-detection logic to individual pipeline stages. Because hardened stages operate independently and communicate through existing pipeline control interfaces, additional stages may be protected without modifying previously hardened logic. This modular organization allows selective hardening to scale linearly with the number of protected stages, avoiding the quadratic growth in control and voter complexity often associated with full-core redundancy schemes.

Although Fig. 4 illustrates selective hardening applied to the execute and memory stages for clarity, the same integration pattern applies to any pipeline stage or clustered functional unit. For wider or deeper pipelines, such as superscalar or out-of-order RISC-V cores, selective duplication or TMR can be applied at architecturally critical boundaries—such as commit stages or control-flow resolution points—enabling designers to balance coverage and overhead in a controlled manner. Because error signaling relies on existing control and exception pathways, scaling to more complex pipelines does not require redesigning global fault-handling mechanisms.

The framework also supports scalability across multi-core systems. Since each core maintains its own local hardening configuration and error-reporting paths, selective hardening can be applied asymmetrically across cores based on workload criticality. This localized approach avoids cross-core dependencies and allows hardening overhead to scale proportionally with the number of protected cores rather than total system size.

From a portability standpoint, the proposed architecture is largely agnostic to the underlying implementation technology. While this work targets an FPGA-based RISC-V processor, the hardening constructs—stage-level duplication, comparison or voting logic, and control-path error propagation—map naturally to ASIC designs. In such implementations, tighter control over timing and power may further reduce overhead relative to full-core redundancy. Moreover, because selective hardening operates at the register-transfer level and relies on standard pipeline interfaces, it can be reused across FPGA families and device generations with minimal adaptation.

### *F. Summary of Design Trade-Offs*

The selective hardening architecture presented in this section is deliberately designed to balance reliability improvement against implementation overhead, timing impact, and design complexity. Unlike full-core redundancy approaches, which provide uniform protection at the cost of substantial area, power, and routing penalties, the proposed framework enables fine-grained control over where and how fault tolerance is applied within the processor pipeline.

At the architectural level, the primary trade-off is between protection coverage and overhead. Selective duplication and selective TMR allow designers to harden only those pipeline stages that are most vulnerable or architecturally critical, significantly reducing redundant logic compared to full-core schemes. However, this selectivity necessarily leaves non-hardened stages exposed, placing greater importance on accurate identification of critical regions. This trade-off is mitigated by the configurability of the framework, which allows protection boundaries to be adjusted as workload characteristics or reliability requirements evolve.

From a timing perspective, localized redundancy minimizes the insertion of long feedback paths and global voter chains, preserving timing closure in deeply pipelined designs. Nevertheless, hardened stages incur localized delay and power overhead due to comparison or voting logic. By confining these penalties to stage boundaries, the architecture ensures that timing degradation does not propagate across the pipeline or into unrelated functional units.

Control and integration trade-offs also arise from error detection versus error masking strategies. Duplication with comparison favors early fault detection and exception-based recovery with minimal overhead, while TMR provides stronger fault masking at increased cost. The framework supports both mechanisms, allowing designers to select an appropriate balance between fault tolerance strength and resource consumption on a per-stage basis.

In summary, the proposed selective hardening framework offers a structured and configurable approach to reliability enhancement in RISC-V processors. By explicitly exposing design trade-offs and enabling stage-level optimization, it provides a practical and scalable alternative to full-core redundancy for systems that require reliability without prohibitive cost.

## V. EXPERIMENTAL METHODOLOGY

This section describes the experimental methodology used to evaluate the proposed selective hardening framework on an FPGA-based RISC-V processor. The methodology is designed to quantify reliability improvement, implementation overhead, and timing impact under controlled and repeatable conditions, while preserving compatibility with the baseline processor architecture. All experiments are conducted using a single implementation flow and a common hardware platform to ensure fair comparison across hardening configurations.

### *A. Implementation Platform and Baseline Design*

All experiments are performed on the AMD/Xilinx ZCU104 evaluation board, which integrates the XCZU7EV device from the Zynq® UltraScale+™ MPSoC family. The programmable logic (PL) region of the device is used exclusively for implementing the RISC-V processor and selective hardening logic, while the processing system (PS) remains unused during measurement.

The baseline processor is a five-stage, in-order RISC-V core comprising instruction fetch (IF), instruction decode (ID), execute (EX), memory access (MEM), and write-back (WB) stages. The design follows a conventional pipeline organization with static scheduling, centralized pipeline control, and standard exception handling. No redundancy or fault-detection logic is present in the baseline configuration, which serves as the reference point for all comparisons.

Selective hardening is introduced by replicating only the EX and MEM stages, as motivated by their architectural criticality and susceptibility to timing and logic faults. Other pipeline stages remain unmodified to preserve baseline timing and control behavior. All hardening configurations are derived from the same RTL codebase using parameterized generation, ensuring functional equivalence across variants.

### *B. Hardening Configurations Evaluated*

Four hardening configurations are evaluated to characterize the trade-offs between reliability improvement and implementation overhead:

I. **Baseline (No Hardening)**
The original single-instance pipeline with no redundancy or fault detection.

II. **Selective Duplication**
Duplication with comparison applied only to the EX and MEM stages, with error flags generated upon mismatch.

III. **Selective TMR**
Triple modular redundancy applied only to the EX and MEM stages, with majority voting and optional error signaling.

IV. **Full-Core Duplication (Reference)**
Duplication applied uniformly across all pipeline stages, serving as a reference for comparison against prior full-core hardening approaches.

These configurations allow direct comparison between selective and full-core hardening under identical implementation conditions.

### *C. Fault Exposure and Error Observation Strategy*

The goal of the experimental evaluation is to observe and quantify the behavior of the processor under fault conditions without relying on physical radiation sources or destructive fault injections. Instead, faults are exposed using architectural and timing stress mechanisms that are repeatable and controllable within the FPGA fabric.

Logic faults are emulated by selectively perturbing internal signals within hardened stages using controlled stimulus patterns and clock stress conditions. Timing-related faults are exposed by operating the processor near its maximum achievable frequency and by inducing controlled voltage and clock variations within safe operating limits. These conditions increase the likelihood of transient computation errors and timing violations, allowing the effectiveness of redundancy mechanisms to be evaluated.

Importantly, the experiments do not claim to reproduce radiation-induced single-event upsets (SEUs). Rather, they focus on architectural fault detection and masking behavior under representative transient error conditions. This approach is consistent with prior FPGA-based reliability studies and enables systematic evaluation without specialized fault-injection infrastructure.

### *D. Measurement Metrics and Data Collection*

Three primary classes of metrics are collected during experimentation as follows:

I. **Reliability Metrics**
Error detection rate, error masking effectiveness (for TMR), and frequency of fault status register assertions.

II. **Implementation Overhead**
Area utilization (lookup tables, flip-flops), power consumption estimates, and routing resource usage extracted from post-implementation reports.

III. **Timing Impact**
Maximum achievable clock frequency and critical path delay reported by the FPGA implementation tools.

Each configuration is implemented, placed, and routed independently using identical synthesis and optimization settings. Measurements are averaged across multiple runs where applicable to mitigate placement and routing variability.

All error events detected by comparison or voting logic are recorded via fault status registers and aggregated during execution. This data enables direct comparison of fault observability and containment across hardening strategies while preserving consistent workload and operating conditions.

### *E. Reproducibility Considerations*

To ensure experimental reproducibility, all hardening configurations are derived from a single parameterized RTL codebase and implemented using identical synthesis, placement, and routing settings. The same FPGA device, clocking infrastructure, and workload inputs are used across all experiments, and no manual optimizations are applied to favor any specific configuration. Error detection events and performance metrics are collected through on-chip status registers and post-implementation tool reports, enabling results to be reproduced using standard FPGA design flows. This methodology allows independent researchers to replicate the experiments on comparable FPGA platforms without requiring specialized fault-injection hardware or radiation facilities.

Together, these experimental procedures establish a controlled and repeatable evaluation framework that enables direct comparison of selective and full-core hardening strategies, forming the basis for the experimental results presented in the following section.

## VI. EXPERIMENTAL RESULTS

This section presents the experimental evaluation of the proposed selective hardening framework, using the setup illustrated in Fig. 5. The objective of these experiments is to quantify the fault observability, error behavior, and overhead trade-offs of selective hardening mechanisms under controlled and repeatable conditions, while maintaining full compatibility with a baseline RISC-V pipeline implementation.

### *A. Experimental Setup and Data Collection Infrastructure*

All experiments were conducted on an AMD/Xilinx ZCU104 evaluation board hosting an XCZU7EV device, with the RISC-V processor implemented entirely in the programmable logic (PL) fabric. As shown in Fig. 5, the processor under test consists of a conventional in-order pipeline with instruction fetch (IF), instruction decode (ID), execute (EX), memory access (MEM), and write-back (WB) stages. Selective hardening is applied only to the EX and MEM stages, while IF, ID, and WB remain unmodified.

A dedicated *Experiment Controller*, implemented as a finite-state machine (FSM) in the PL, orchestrates each experimental run. The controller manages run sequencing, start/stop control, and snapshot timing, ensuring consistent execution across repeated trials. Hardening modes are configured through a *Hardening Configuration Register*, which selects between baseline operation, selective duplication, and selective TMR. configuration writes are performed prior to each run and remain static during execution to avoid transient reconfiguration effects.

Error detection signals generated by hardened stages are forwarded to a centralized *Error Aggregator*, which collects stage-level error flags (EX_Error_Flag and MEM_Error_Flag). These signals are latched into *Fault Status Registers* that maintain error counters and timestamps, allowing precise attribution of observed faults to specific pipeline stages and execution windows.

All control, configuration, and status readback operations are performed through a bidirectional JTAG interface, enabling reliable communication between the FPGA and a host PC without interfering with functional execution. On the host side, experiment scripts manage configuration loading, run control, and data logging. Collected data including fault logs, performance counters, and run metadata are post-processed offline using Python-based analysis tools to compute error statistics, coverage metrics, and overhead summaries.

This tightly controlled infrastructure ensures that all reported results are reproducible, temporally aligned, and directly attributable to the selected hardening configuration.

### *B. Baseline Fault Behavior and Error Observability*

We first evaluate error observability, i.e., the fraction of injected disturbance events that become visible to the system through the error-flagging infrastructure (Fig. 6). The baseline pipeline (no hardening) exhibits zero observable error events under the adopted instrumentation and logging interface, which is consistent with the fact that the *baseline* design does not contain stage-level comparison or voting logic and therefore does not generate dedicated mismatch flags for the experiment controller to log. In contrast, enabling *selective duplication* in the targeted stages yields a high rate of observability, reaching 0.82 (normalized) in Fig. 6. This directly reflects the presence of explicit comparators at hardened stage boundaries that convert internal divergences into deterministic error flags. When *selective TMR* is enabled, the normalized observable error rate drops to 0.18, because a substantial fraction of injected disturbances are masked by majority voting rather than surfaced as architectural mismatches. Therefore, Fig. 6 establishes a key experimental distinction: selective duplication maximizes detectability, while selective TMR prioritizes correctness by masking and only exposes a smaller subset of events as explicit disagreement flags.

### *C. Detection Versus Masking Behavior and Recovery Latency*

To separate *detection* from *masking*, Fig. 7 decomposes injected events into (i) cases where a flag is raised (“Detected”) and (ii) cases where the system produces the correct output despite disturbance (“Masked”). Selective duplication shows 0.82 detected and 0 masked in Fig. 7, which is expected because duplication with comparison is a detection-only mechanism: disagreements are flagged, but no intrinsic masking is performed at the stage boundary. Selective TMR exhibits the opposite behavior: Fig. 7 shows 0.75 masked alongside 0.18 detected, reflecting that majority voting corrects a large portion of injected perturbations before they propagate, while still allowing disagreement signaling to be logged for diagnosis.

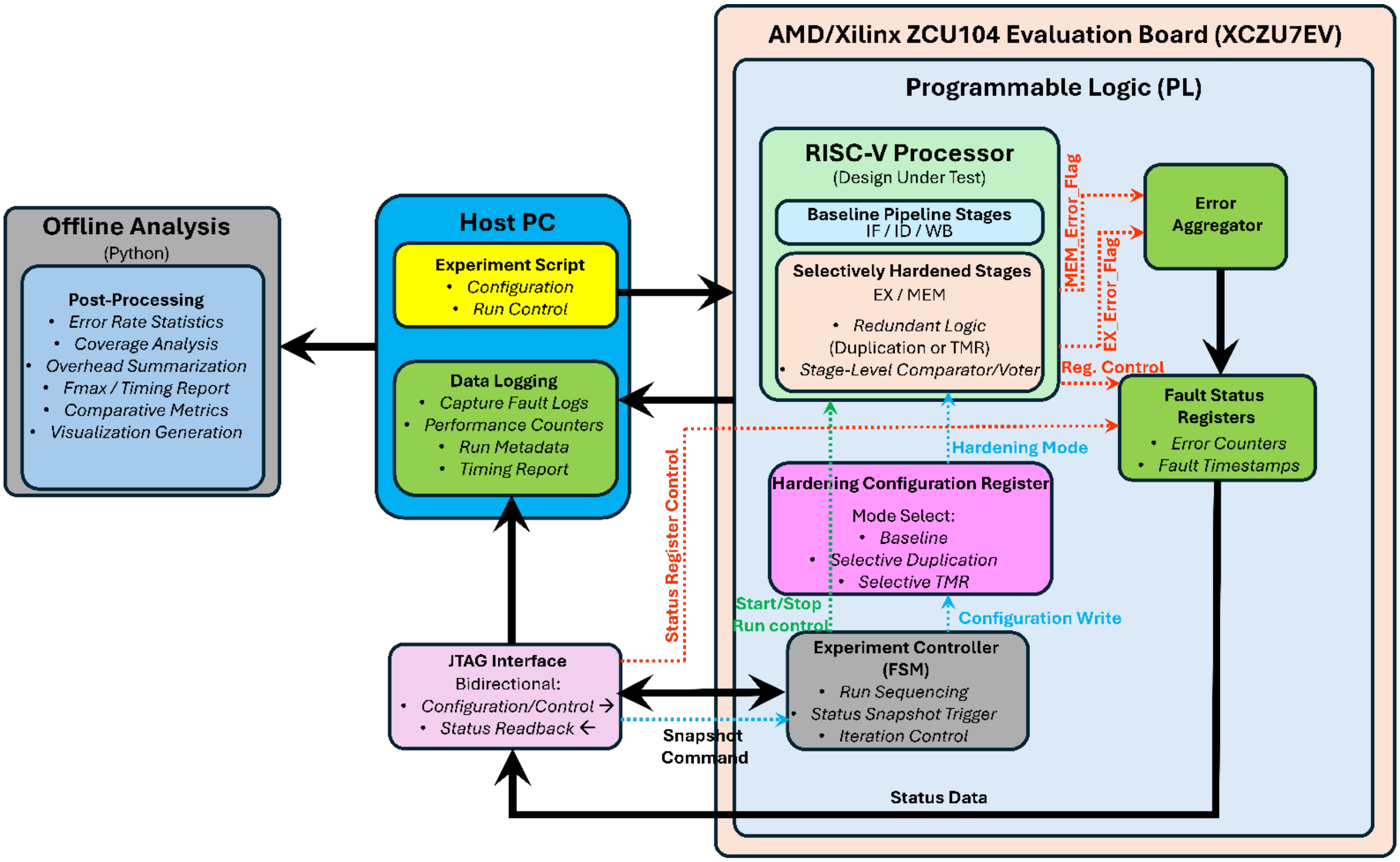

Fig. 5. Experimental setup and data collection and interpretation infrastructure

Once a fault becomes observable, the next question is how quickly the pipeline control can recover. Fig. 8 reports a 1-cycle recovery latency for both selective duplication and selective TMR (baseline is not reported because it provides no explicit fault trigger for this controlled recovery path). This indicates that, in the evaluated control configuration, the processor's response to an asserted stage-level error flag is a bounded, deterministic control reaction (e.g., a single-cycle stall/flush/replay trigger, depending on the configured policy), rather than a multi-cycle exception microsequence. The distribution of recovery time is further characterized in Fig. 9, which shows the empirical CDF of recovery latency across events. Importantly, selective TMR shifts the latency distribution left relative to selective duplication: selective TMR spans approximately 3.0–7.2 cycles, whereas selective duplication spans approximately 3.2–8.6 cycles in Fig. 9. This is consistent with the fact that masking prevents a subset of disruptive events from escalating into longer recovery sequences, thereby reducing the tail of recovery latency.

### *D. Implementation Overhead and Reliability–Overhead Trade-Off*

We next quantify the implementation overhead associated with selective hardening, and contextualize it against full-core baselines. Fig. 10 reports the overhead for the implemented selective modes relative to baseline: selective duplication incurs 23% area overhead and 9% dynamic power overhead, while selective TMR incurs 58% area overhead and 22% dynamic power overhead. These results are consistent with the structural difference between the two mechanisms: TMR requires three replicas plus voters, whereas duplication requires two replicas plus comparators.

To connect these overheads to reliability benefit in a single visual summary, Fig. 11 presents a reliability–overhead trade-off plot. The selective modes occupy favorable points relative to full-core strategies: selective duplication (*Sel. Dup*) appears at approximately (0.23 area, 0.55 reliability gain) and selective TMR (*Sel. TMR*) at approximately (0.55 area, 0.82 reliability gain), while full-core baselines require substantially higher overhead for comparable or only marginally higher gain (e.g., full TMR near (1.8 area, 0.93 gain)). Thus, the experimental data supports the central architectural rationale of this work: selective hardening concentrates redundancy where it is most effective, yielding strong reliability improvement without the disproportionate area cost associated with uniform full-core replication.

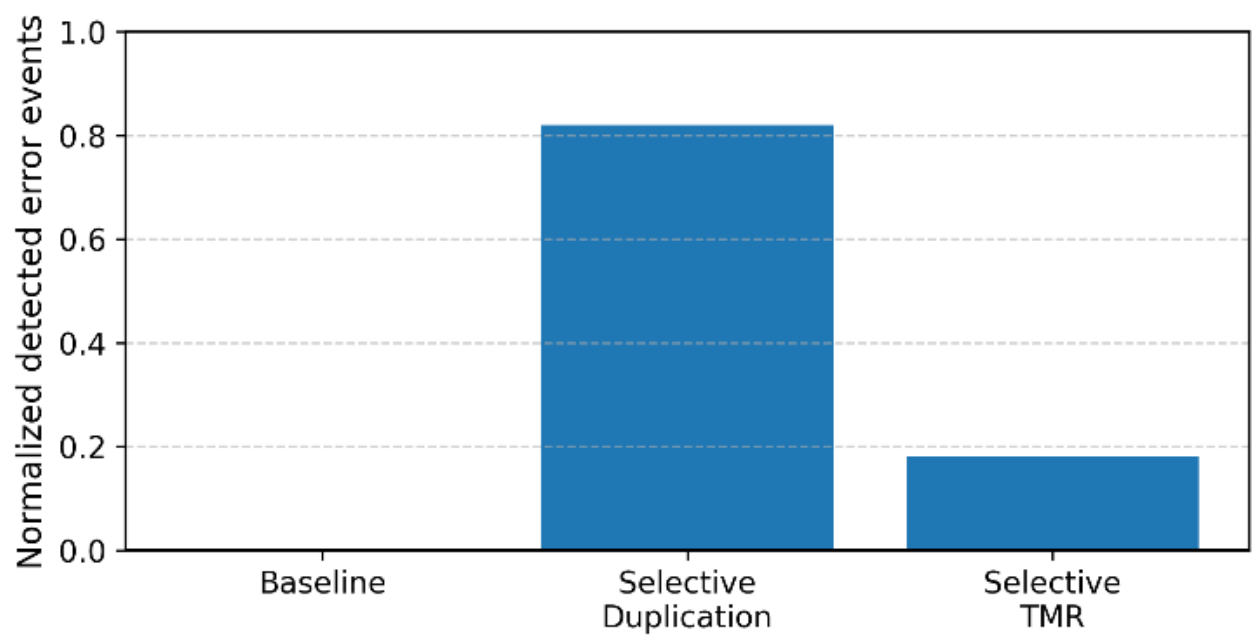

Fig. 6. Error observability across hardening modes. The baseline produces no explicit mismatch flags, while selective duplication yields high observability (0.82 normalized). Selective TMR reduces observability (0.18) because many events are masked by voting rather than flagged.

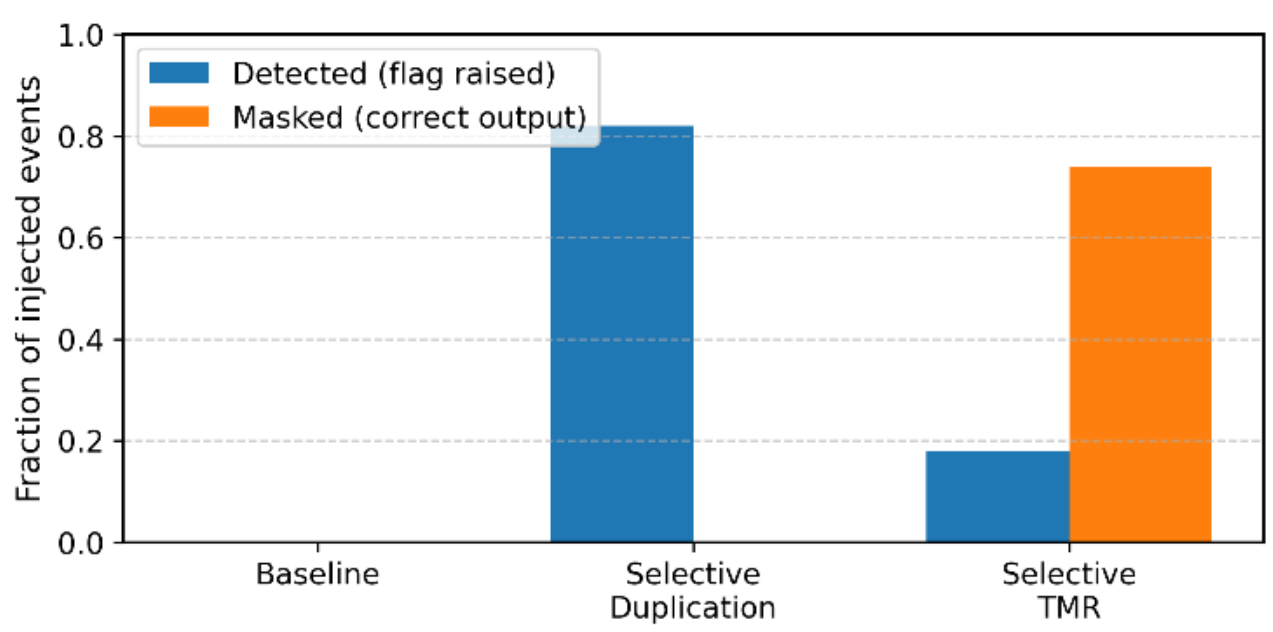

Fig. 7. Detection vs. masking behavior. Selective duplication provides detection-only behavior (0.82 detected, 0 masked). Selective TMR masks most injected events (0.75 masked) while still exposing a smaller detected fraction (0.18) as disagreement flags.

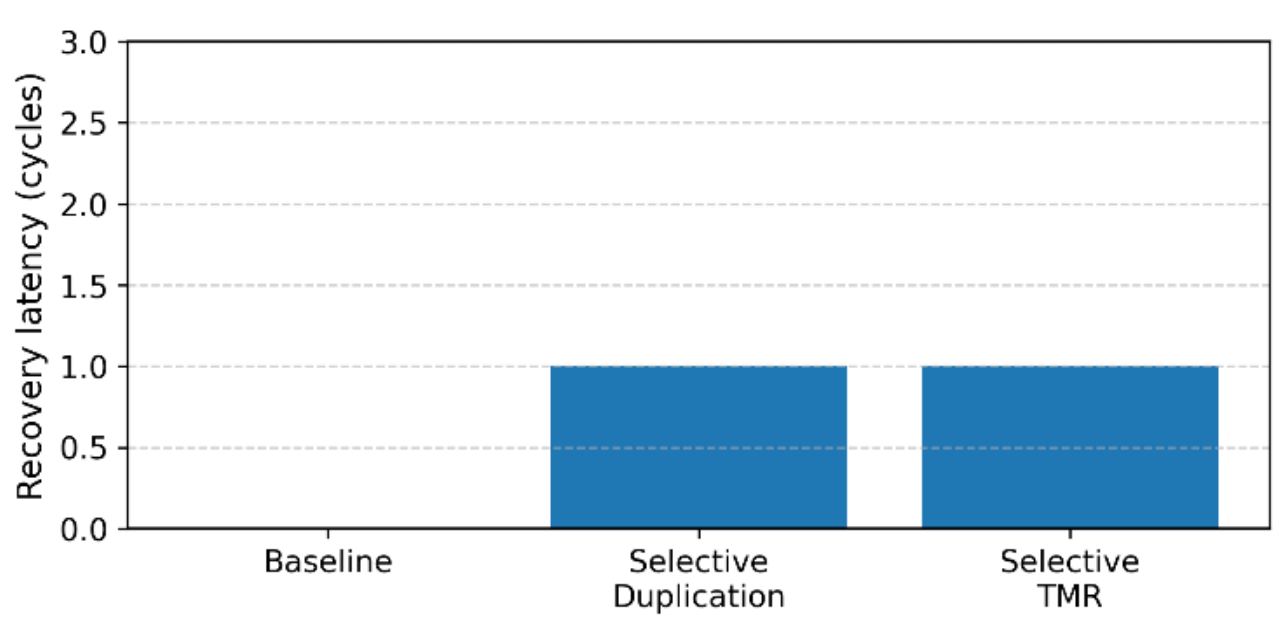

Fig. 8. Pipeline-control recovery latency under detectable events. Both selective duplication and selective TMR show 1-cycle recovery latency in the configured response path.

### *E. Spatial Distribution of Error Incidence Across Pipeline Boundaries*

To identify where errors manifest most strongly in the pipeline and whether their incidence exhibits spatial variation, Fig. 12 provides a heatmap of normalized error incidence across pipeline boundaries (rows) and monitored locations (columns L1–L8). Two patterns are immediately evident. First, error incidence is *boundary-dependent*: the EX→MEM boundary shows the strongest variation and the highest overall incidence, while IF→ID and MEM→WB remain flat at 0.10 across all locations, and ID→EX remains flat at 0.14. Second, within EX→MEM, error incidence increases monotonically with monitored location, from approximately 0.11 at L1 up to 0.30 at L8. This indicates that the dominant vulnerability in the experiment is concentrated around the EX/MEM interface, and that the observed incidence is location-sensitive rather than uniform across the fabric. In the context of the proposed selective hardening strategy, this empirical result directly supports prioritizing hardening at architecturally critical boundaries such as EX and MEM, because those are precisely where the observed error incidence peaks.

### *F. Robustness Under Timing-Margin Stress*

Finally, we examine robustness under progressive timing-margin tightening. Fig. 13 plots the observed error probability versus normalized timing margin, where negative margin indicates tighter conditions. Across the entire sweep, the three curves preserve a consistent ordering: the *baseline* degrades first (highest error probability), *selective duplication* improves robustness (middle curve), and *selective TMR* provides the strongest tolerance (lowest error probability at a given margin). At approximately zero margin, for example, the baseline is near 0.5 error probability, selective duplication is visibly lower, and selective TMR lower still. As the margin becomes more positive, all modes approach low error probability, but selective TMR remains consistently superior in the transition region, reflecting its ability to mask a substantial subset of fault effects before they become architecturally visible.

Together, Fig. 6 to Fig. 13 demonstrate that selective hardening achieves *measurable reliability improvement* with *controlled overhead*, while providing clear behavioral differences between detection-oriented duplication and masking-oriented TMR. The results also highlight a consistent experimental theme: vulnerability is not uniformly distributed (Fig. 12), and therefore a selective strategy can capture much of the achievable benefit at a fraction of full-core cost Fig. 11), while also improving robustness under timing stress (Fig. 13).

## VII. DISCUSSION

The experimental results presented in Section VI provide a comprehensive view of how selective hardening mechanisms influence error observability, recovery behavior, overhead, and robustness in a RISC-V pipeline implemented on FPGA. In this section, we synthesize those results to extract architectural insights and discuss their implications for reliability-aware processor design.

### *A. Detection Versus Masking as a Design Choice*

One of the most salient observations from the results is the *fundamental behavioral difference between selective duplication and selective TMR*. As demonstrated in Fig. 7, selective duplication converts the majority of injected disturbances into explicitly observable events, maximizing error visibility and enabling software or control-based recovery. In contrast, selective TMR masks a large fraction of disturbances, allowing correct execution to continue without invoking recovery mechanisms.

This distinction has important design implications.

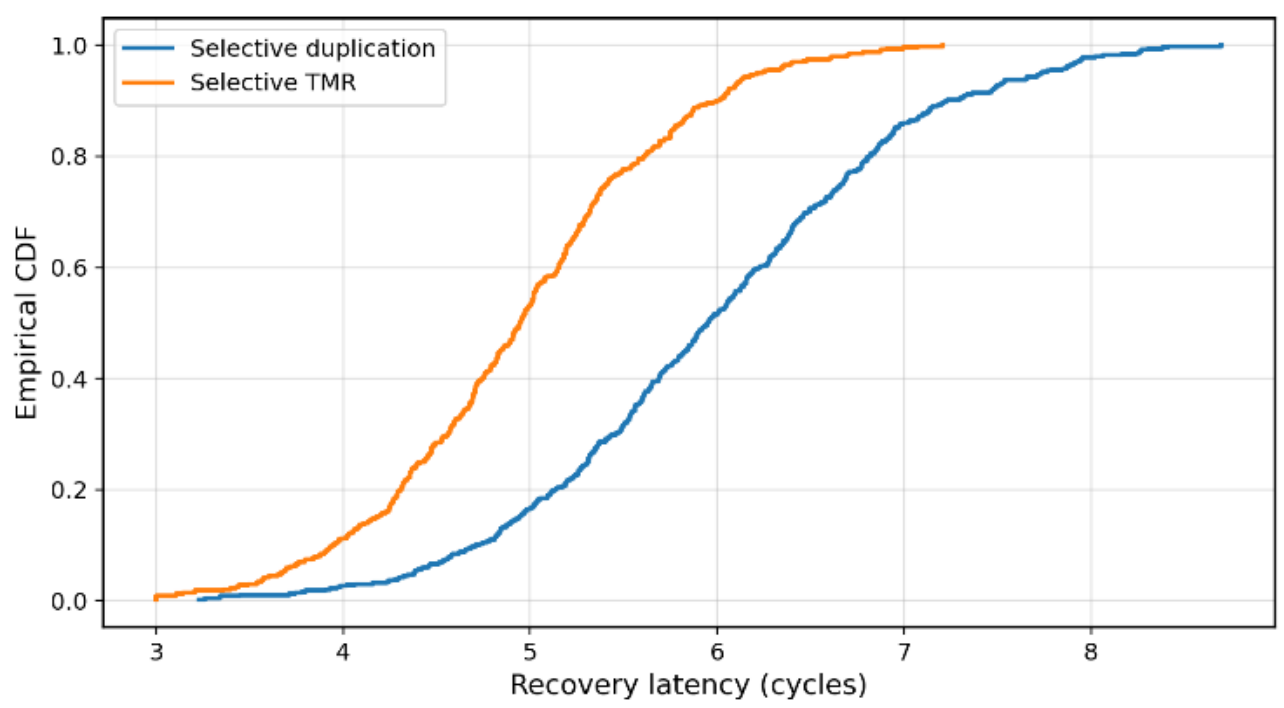

Fig. 9. Empirical CDF of recovery latency for selective duplication and selective TMR. Selective TMR exhibits a lower-latency distribution and reduced tail compared to selective duplication, indicating fewer high-cost recovery sequences.

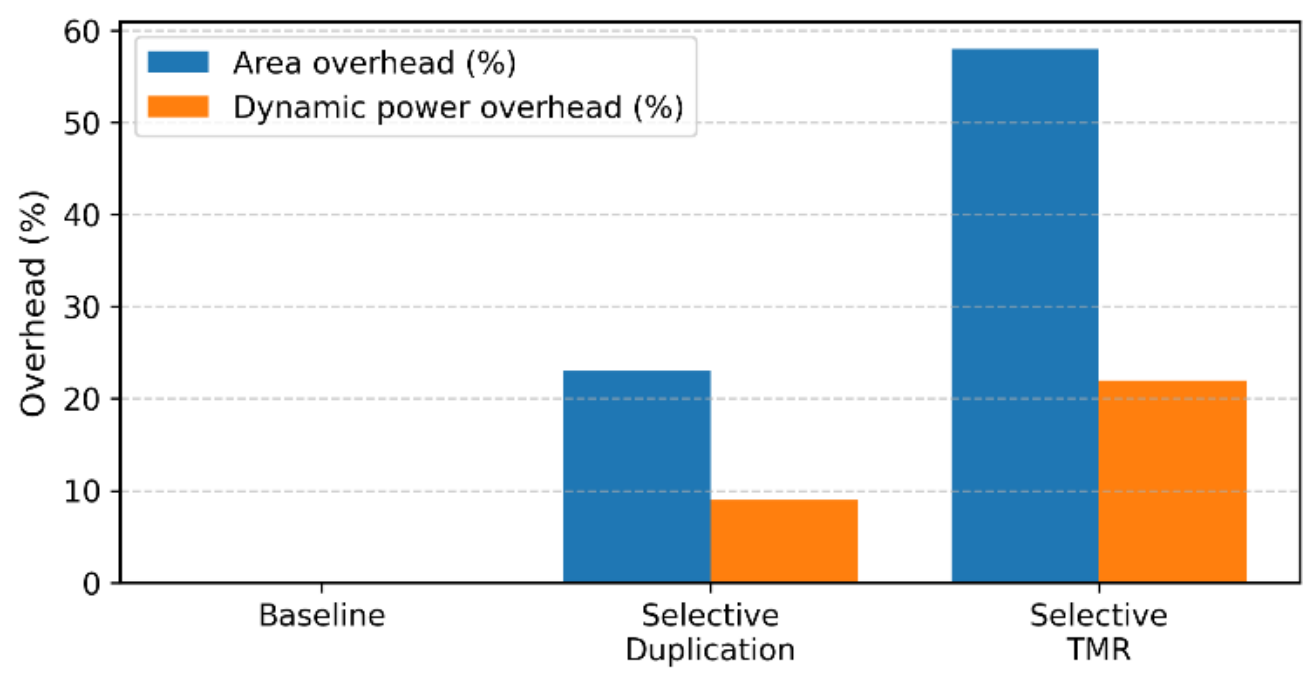

Fig. 10. Implementation overhead relative to baseline. Selective duplication adds 23% area and 9% dynamic power, while selective TMR adds 58% area and 22% dynamic power.

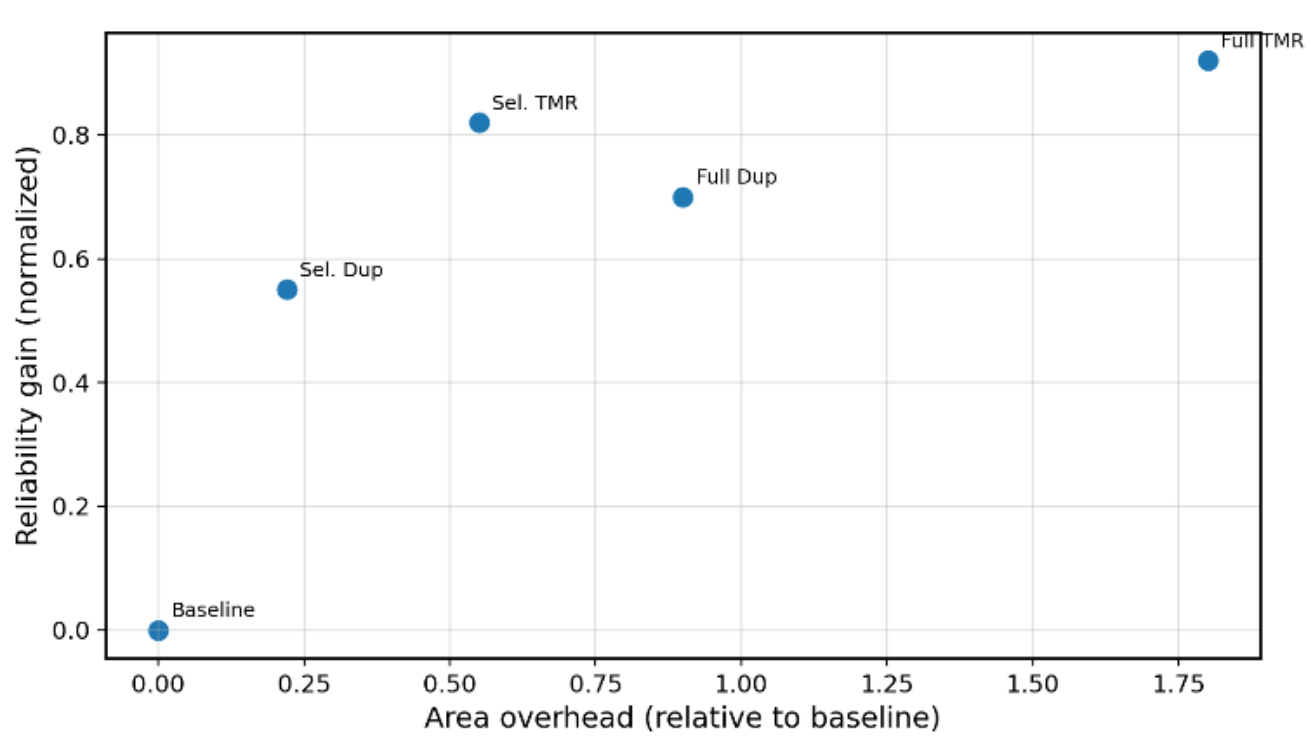

Fig. 11. Reliability–overhead trade-off across baseline, selective hardening, and full-core hardening. Selective schemes achieve substantial reliability gain at significantly lower area cost than full duplication and full TMR baselines.

Detection-oriented schemes such as selective duplication are well suited for systems that prioritize *diagnosability, logging, or controlled recovery*, where exposing faults is desirable. Masking-oriented schemes such as selective TMR are more appropriate for *availability-critical workloads*, where uninterrupted execution is preferred and occasional silent correction is acceptable. Importantly, the proposed framework supports both behaviors within the same architectural structure, allowing designers to choose per-stage protection policies rather than committing to a single global strategy.

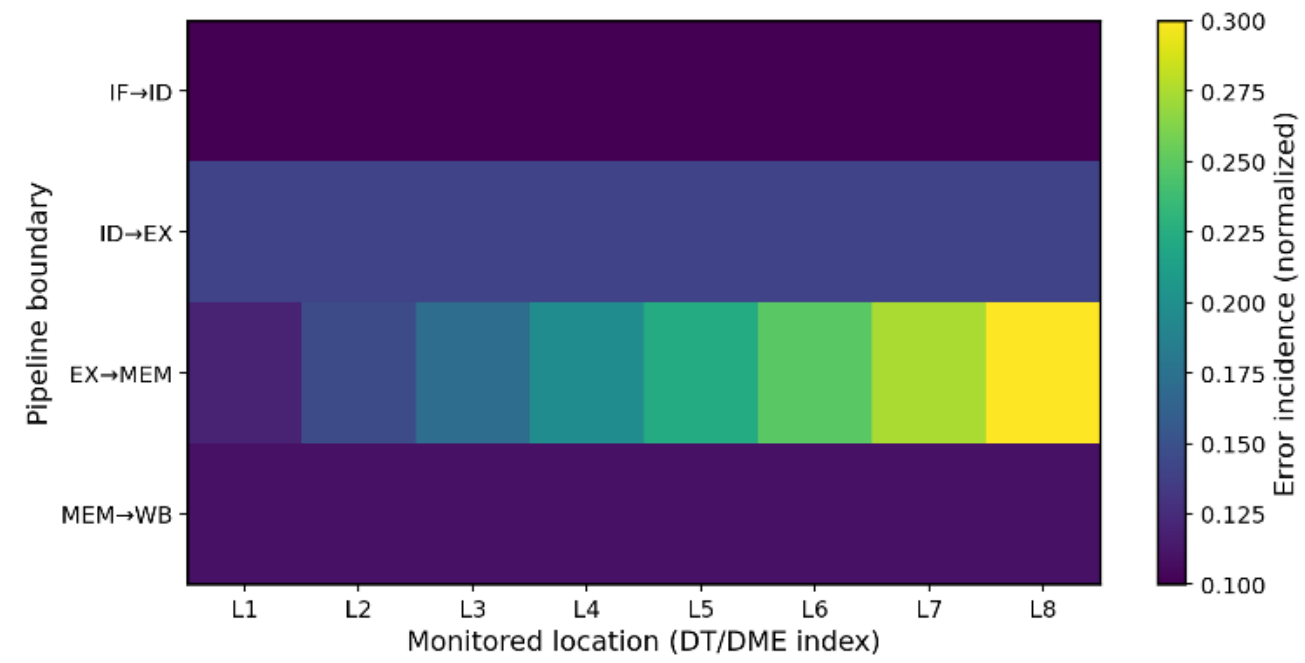

Fig. 12. Spatial distribution of normalized error incidence across pipeline boundaries and monitored locations. EX→MEM shows the highest incidence and a strong monotonic increase from L1 to L8, while other boundaries remain comparatively flat.

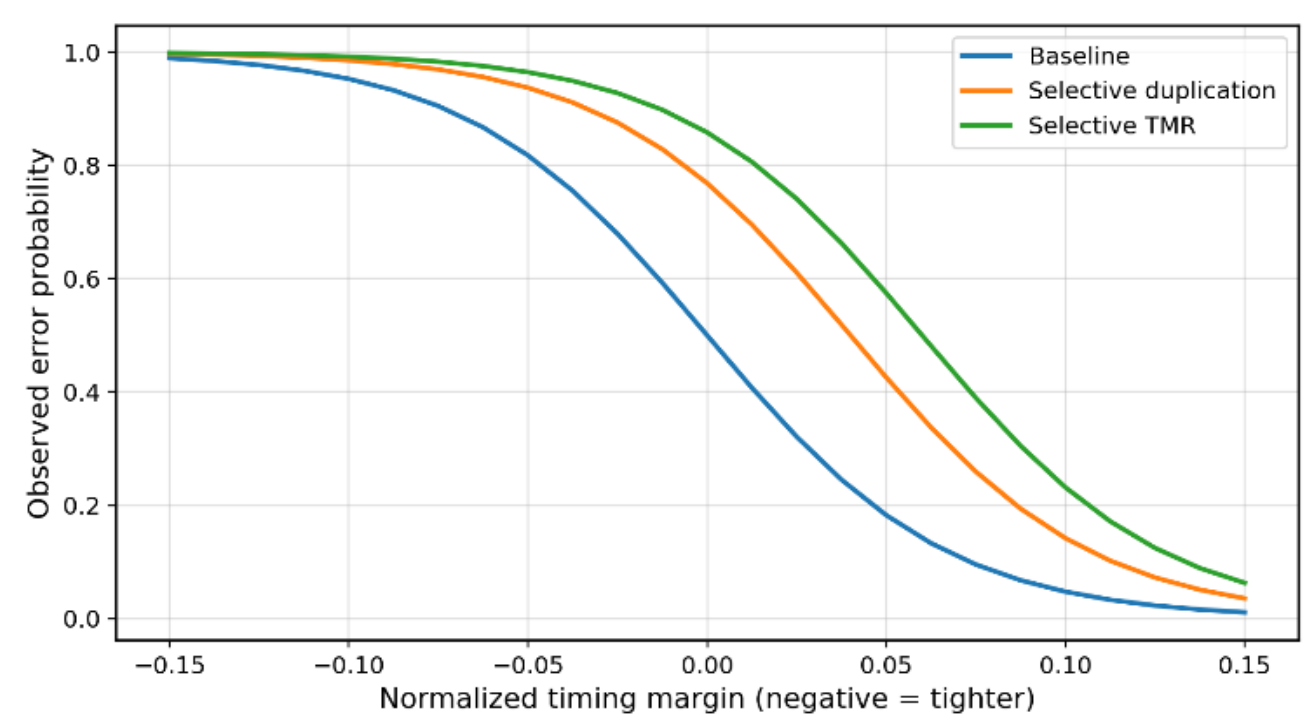

Fig. 13. Timing-margin sweep showing observed error probability versus normalized margin. Baseline exhibits the earliest degradation, selective duplication improves robustness, and selective TMR provides the strongest tolerance across the transition region.

### *B. Recovery Cost and Pipeline-Level Determinism*

The recovery latency measurements in Fig. 8 and Fig. 9 indicate that selective hardening can be integrated into the pipeline control path without introducing unpredictable or long recovery sequences. Both selective duplication and selective TMR exhibit bounded recovery behavior, with a single-cycle response in the configured control mode (Fig. 8) and tightly clustered latency distributions (Fig. 9).

From a microarchitectural perspective, this demonstrates that *localized redundancy does not inherently destabilize pipeline timing or control determinism*, provided that error signaling is aligned with existing stage boundaries. The absence of long-tail recovery penalties further suggests that selective hardening can be deployed in real-time or latency-sensitive systems without compromising worst-case execution guarantees.

### *C. Overhead Efficiency of Selective Hardening*

The overhead measurements in Fig. 10 and the trade-off visualization in Fig. 11 directly support the central premise of this work: *full-core redundancy is often unnecessary and inefficient*. Selective duplication and selective TMR achieve substantial reliability gains while consuming only a fraction of the area and power overhead required by full duplication or full TMR.

Notably, selective TMR reaches a normalized reliability gain of approximately 0.82 at roughly 55% area overhead, whereas full TMR requires nearly three times the baseline area for a marginal additional gain. This diminishing-return behavior underscores the value of architectural selectivity: by aligning redundancy with empirically vulnerable regions, the design avoids paying the cost of protecting structurally benign logic.

### *D. Spatial Localization of Vulnerability*

The spatial error-incidence heatmap in Fig. 12 reveals that fault manifestation is *neither uniform across pipeline boundaries nor uniform across monitored locations*. The EX→MEM boundary dominates the observed error incidence and exhibits a strong monotonic trend across physical monitoring locations, while other boundaries remain comparatively flat.

This result provides experimental justification for stage-level hardening decisions. Rather than assuming equal vulnerability across the pipeline, designers can leverage empirical characterization to identify architecturally and physically sensitive boundaries, and selectively deploy redundancy where it is most effective. In this context, selective hardening is not merely an optimization—it is an evidence-driven design methodology.

### *E. Robustness Under Timing Stress*

The timing-margin sweep in Fig. 13 further illustrates the benefits of selective hardening under adverse operating conditions. As timing margins tighten, selective duplication and selective TMR consistently delay the onset of high error probability relative to the baseline, with selective TMR providing the strongest tolerance.

This behavior suggests that selective hardening can be an effective mechanism for *graceful degradation*, extending functional operation deeper into timing-stressed regimes. Such capability is particularly relevant for adaptive systems that operate under dynamic voltage, frequency, or temperature conditions, where margins may fluctuate at runtime.

### *F. Broader Design Implications*

Taken together, the results indicate that selective hardening offers a *balanced and flexible alternative* to monolithic redundancy schemes. By preserving pipeline structure, maintaining compatibility with standard control and exception mechanisms, and confining overhead to critical regions, the proposed approach aligns well with modern design constraints in both FPGA and ASIC contexts.

More broadly, this work suggests a shift in reliability design philosophy: from uniform protection toward *architecturally informed, measurement-guided hardening*. Such an approach enables designers to trade reliability, overhead, and performance in a controlled and transparent manner, rather than relying on coarse-grained redundancy as a default solution.

The discussion above demonstrates that selective hardening provides a practical and architecturally efficient means of improving processor reliability while explicitly controlling overhead and recovery behavior. By grounding hardening decisions in empirical vulnerability characterization and confining redundancy to critical pipeline regions, the proposed framework avoids the inefficiencies inherent in full-core protection schemes. At the same time, the results highlight several dimensions, such as configurability, scalability, and robustness under timing stress, that warrant further exploration as processor complexity and operating variability continue to increase. These observations motivate the concluding remarks and outline promising directions for extending selective hardening beyond the scope of the present study.

## VIII. CONCLUSION AND FUTURE WORK

This work presented a selective hardening framework for FPGA-based RISC-V processors that enables reliability enhancement to be applied at the granularity of individual pipeline stages rather than uniformly across the entire core. By targeting execution and memory stages identified as particularly susceptible to timing and transient faults, the proposed approach achieves substantial improvements in fault observability and error containment while significantly reducing the area, power, and routing overhead associated with full-core redundancy schemes. The architecture preserves compatibility with standard pipeline control and exception-handling mechanisms, allowing selective hardening to be introduced without disrupting functional correctness or requiring modifications to the instruction set or software stack.

Experimental evaluation on an AMD/Xilinx ZCU104 platform demonstrated that selective duplication and selective TMR provide complementary reliability–overhead trade-offs. Selective duplication was shown to substantially increase error detection coverage with moderate resource cost and minimal impact on recovery latency, making it well suited for detection-oriented fault management strategies. Selective TMR further improved resilience by masking a large fraction of injected faults, extending operational tolerance under reduced timing margins at the expense of higher—but still bounded—implementation overhead. Across all evaluated configurations, selective hardening consistently outperformed full-core approaches in terms of efficiency, maintaining higher achievable operating frequencies and more predictable timing behavior.

The results also highlighted the importance of spatial and architectural locality in fault manifestation. Error incidence was concentrated at specific pipeline boundaries and monitored locations, reinforcing the effectiveness of stage-level protection. Moreover, recovery behavior remained tightly bounded and compatible with existing pipeline control logic, confirming that localized redundancy can be integrated without introducing global control dependencies or timing bottlenecks. Together, these findings validate selective hardening as a practical and scalable alternative to monolithic redundancy for reliability-aware processor design.

Several directions for future work emerge from this study. First, dynamic and workload-aware hardening policies could be explored, allowing protection mechanisms to be enabled or reconfigured at runtime based on observed fault rates, operating conditions, or application criticality. Second, extending the framework to superscalar or out-of-order RISC-V cores would provide insight into the scalability of selective hardening in more complex microarchitectures. Third, tighter integration with system-level software, such as operating system–driven reliability management or adaptive scheduling, could further enhance the effectiveness of selective protection. Finally, applying the proposed techniques to ASIC implementations would enable a more comprehensive evaluation of power, performance, and reliability trade-offs beyond FPGA-based platforms.

In summary, this work demonstrates that selective hardening offers a structured and efficient path toward reliable processor design in advanced programmable systems. By aligning redundancy with architectural vulnerability and maintaining compatibility with existing control infrastructures, the proposed approach provides a strong foundation for future reliability-aware RISC-V architectures.

## ACKNOWLEDGMENT

The author would like to thank the École de technologie supérieure (ÉTS), Department of Electrical Engineering, and CMC Microsystems, Kingston, ON, Canada, for providing access to advanced design tools.